\documentclass[epsfig,superscriptaddress,preprint]{revtex4}
\usepackage{amsmath}
\usepackage{amsfonts}
\usepackage{amssymb}
\usepackage{graphicx}
\usepackage{graphicx}% Include figure files
\usepackage{dcolumn}% Align table columns on decimal point
\usepackage{bm}% bold math
\usepackage[sort&compress]{natbib}
\bibpunct{[}{]}{,}{n}{,}{,}

\begin{document}

\title{Observation of beam loading in a laser-plasma accelerator}% Force line breaks with \\

\author{C. Rechatin}

\affiliation{%
Laboratoire d'Optique Appliqu\'ee, ENSTA, CNRS, Ecole Polytechnique, UMR 7639, 91761 Palaiseau, France\\
}%
\author{X. Davoine}
\affiliation{%
CEA, DAM, DIF, Bruy\`eres-le-Ch\^atel, 91297 Arpajon, France}
\author{A. Lifschitz}
\affiliation{%
Laboratoire d'Optique Appliqu\'ee, ENSTA, CNRS, Ecole Polytechnique, UMR 7639, 91761 Palaiseau, France\\
}%
\affiliation{Laboratoire de Physique des Gaz et des Plasmas, CNRS, UMR 8578, Universit\'e Paris XI, B\^atiment 210, 91405 Orsay cedex, France\
}
\author{A. Ben Ismail}
\affiliation{%
Laboratoire d'Optique Appliqu\'ee, ENSTA, CNRS, Ecole Polytechnique, UMR 7639, 91761 Palaiseau, France\\
}%
\author{J. Lim}
\affiliation{%
Laboratoire d'Optique Appliqu\'ee, ENSTA, CNRS, Ecole Polytechnique, UMR 7639, 91761 Palaiseau, France\\
}%
\author{E. Lefebvre}
\affiliation{%
CEA, DAM, DIF, Bruy\`eres-le-Ch\^atel, 91297 Arpajon, France}
\author{J. Faure}
\affiliation{%
Laboratoire d'Optique Appliqu\'ee, ENSTA, CNRS, Ecole Polytechnique, UMR 7639, 91761 Palaiseau, France\\
}%
\author{V. Malka}
\affiliation{%
Laboratoire d'Optique Appliqu\'ee, ENSTA, CNRS, Ecole Polytechnique, UMR 7639, 91761 Palaiseau, France\\
}%}
\date{\today}

\begin{abstract}

Beam loading is the phenomenon which limits the charge and the beam
quality in plasma based accelerators. An experimental study conducted
with a laser-plasma accelerator is presented. Beam loading manifests
itself through the decrease of the beam energy, the reduction of dark
current and the increase of the energy spread for large beam
charge. 3D PIC simulations are compared to the experimental results
and confirm the effects of beam loading. It is found that, in our
experimental conditions, the trapped electron beams generate
decelerating fields on the order of 1 GV/m/pC and that beam loading
effects are optimized for trapped charges of about 20 pC. 

\end{abstract}
\maketitle

The concept of laser wakefield accelerator, as first developed by
Tajima and Dawson \cite{TajimaPhysRevLett.43.267} relies on the
excitation of a longitudinal plasma wave by the ponderomotive
force of a laser pulse. The driven electric field, exceeding hundreds
of GV/m, can be used to accelerate electrons to relativistic energies
in a millimeter scale. Over the past few years, this compact acceleration
technique has made remarkable progress, producing quasi mono-energetic
electron bunches at the 100 MeV level \cite{2004Nature1,2004Nature2,2004Nature3}
and then up to the GeV \cite{LBNL1GeV}, as well as increasing
stability and tunability \cite{NatureCPI}.

A fundamental limit of these accelerators is due to the field perturbation
driven by the accelerated electron bunch itself. Indeed, when a bunch
of electrons is accelerated, it drives a plasma oscillation which
can cancel out the laser wakefield. This phenomenon, known as ``beam
loading'', ultimately limits the charge that
can be accelerated since for a given charge, the longitudinal field
will no longer be accelerating over the whole bunch length.
Before reaching this limit, beam loading also impacts the beam quality
since the trailing electrons of the bunch witness the superposition of
the  laser  wakefield  and  the  plasma wave  driven  by  the  leading
electrons of the bunch. Therefore, when the bunch is carefully shaped,
the total electric  field can be made constant  over the bunch length,
which          minimizes           the          energy          spread
\cite{kats87,Reitsma,Tzoufras}. However,  when uncontrolled, this fast
varying beam loading field might  lead to an undesirable growth of the
energy  spread.  This effect  is  therefore  of  major importance  for
designing the next generation  of laser plasma accelerators delivering
a high quality electron source. 

In this letter we give, to our knowledge, the first experimental
observation of beam loading in a laser-plasma accelerator. To
obtain those conclusive evidences, we have used an optical scheme to
control the injection of electrons \cite{Esarey}.
In this scheme, electrons gain momentum in the ponderomotive beatwave
created by the collision of the main laser pulse (pump pulse) with a
second laser pulse (injection pulse) \cite{NatureCPI,Fubiani}, and therefore
have enough energy to be trapped in the wakefield. This injection 
mechanism has proven to inject electrons in a stable and reproducible
manner \cite{NatureCPI,PPCF}. Moreover, by 
decoupling the injection and acceleration processes, it is possible to
gain control over electron beam parameters by only changing the
injection pulse parameters \cite{PRLrec}. It therefore enables us to
load various charges in the plasma wakefield without changing the
laser driving the plasma wave, nor the plasma parameters.

The experiment was conducted with the LOA ``Salle jaune'' Ti:Sa laser system, 
that delivers two linearly polarized pulses of 30 fs. The pump pulse
is focused to an intensity of $I_0=4.6\times 10^{18}\mbox{W.cm}^{-2}$,
giving a normalized amplitude of $a_0=1.5$. The injection pulse is
focused to a maximal intensity of $I_1=4\times
10^{17}\mbox{W.cm}^{-2}$, giving a normalized amplitude of $a_1=0.4$. A
half-wave plate followed by a polarizer enables us to continuously
reduce the injection pulse intensity. We used a 3 mm supersonic
Helium gas nozzle with a 2.1 mm well defined density plateau at
electron density $n_e=5.7\;\times 10^{18}\mbox{cm}^{-3}$; the gas
was fully ionized early in the interaction. Electrons are injected
at the collision of the two laser pulses colliding at an angle of $176^{\circ}$, either $400\;
\mu\mbox{m}$ or $250\;\mu\mbox{m}$ before the center of the nozzle, and are
accelerated over the remaining length of plasma i.e. 1.45 mm and 1.3 mm
respectively. They are then deflected by a dipole magnet of 1.1 T over 10 cm
before hitting a LANEX screen, which gives access to the spectral
information of the electron bunch above 45 MeV \cite{aimantyannick}.

As stated before, beam loading can manifest itself through a
correlation between the bunch energy spread and charge. In our experiments, we have
changed the loaded charge by changing $a_1$, which also affects the
initial volume of electrons in phase space (injection volume) and thus directly leads to a change of energy
spread \cite{PRLrec}. Disentanglement of the two processes would require
extensive use of simulations for which the energy spread is
unfortunately the observable less robust to initial parameters fluctuations.
Thus,  in  this  paper,  we   have  concentrated  on  other  means  to
experimentally observe beam loading.

Beam  loading  reduces the  energy  of  the  trailing
electrons of the bunch since they experience the field perturbation of
the  leading electrons.   Therefore,  through beam  loading, the  mean
energy and peak energy of the electron bunch should be correlated with the load.

\begin{figure}
\centering
\includegraphics[width=0.8\linewidth]{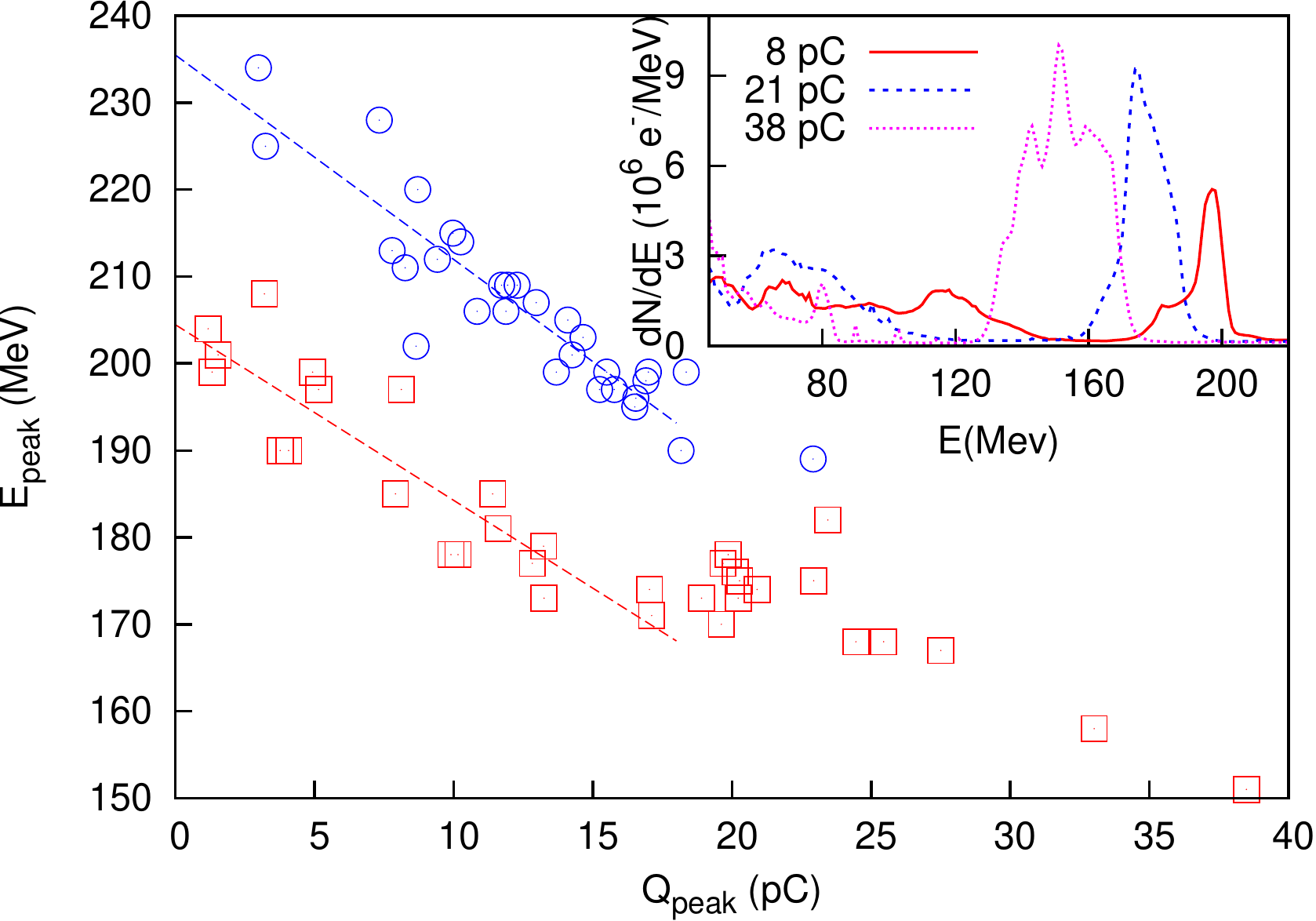}
\caption{Experimental results: Bunch peak energy versus charge trapped in the peak. Red squares correspond to the
 varying $a_1$ data set, blue circles to 30 consecutive shots with
 laser fluctuations. Inset: electron spectra obtained for three
 different injection laser amplitude, from left to right
 $a_1=0.4$ (38 pC), 0.24 (21 pC) and 0.1 (8 pC).
\label{expen} }
\end{figure}

The inset of Fig.\ref{expen} represents the electron spectra obtained
with three different injection amplitudes. It clearly shows a decrease
of energy with increasing injection amplitude and beam charge. Here,
the peak energy of the quasi mono-energetic component goes from
197 MeV, for an injected charge of 8 pC ($a_1=0.1$), to 151 MeV, for a
charge of 38 pC ($a_1=0.4$). 
To represent more data and give statistically clearer results, we
represent in Fig.\ref{expen} the peak-energy versus the charge
of the quasi mono-energetic peak for two complete data sets. One set of
data (circles) is obtained by looking at the fluctuations of charge and spectra
over 30 shots, for which the only variations are the laser intensity
and pointing fluctuations at the collision position of
$400\;\mu\mbox{m}$ before the center of the nozzle
($z_{coll}=-400\;\mu\mbox{m}$). The other set of data (squares) is
obtained for injection at $z_{coll}=-250\;\mu\mbox{m}$ by varying $a_1$,
 and thus forcing a change of trapped charge in the first bucket over
 a wider  range. Those curves  clearly confirm the  strong correlation
 between trapped charge and 
energy. These  data points also  exhibit, for small charges,  a linear
slope  (dashed lines).  When  normalized by  the acceleration  length,
those   slopes   are   similar:   1.6   GV/m/pC   for   injection   at
$z_{coll}=-400\;\mu\mbox{m}$  (circles),1.55 GV/m/pC for  injection at
$z_{coll}=-250\;\mu\mbox{m}$ (squares).

The inset of Fig.\ref{expen} also gives a typical evolution of the
energy spread. For small loads, the energy spread stays small and in
this case, close to the spectrometer resolution (5 \%). For higher loads,
here above 25 pC, the energy spread grows fast and substructures appear in
the quasi mono-energetic component of the spectrum.

Another possible way to diagnose beam loading is to monitor
the charge trapped in the trailing plasma buckets. Those electrons, also
heated in the beat-wave, are not trapped in the first bucket mainly
because of  wakefield inhibition  \cite{poprechatin}, but they  can be
trapped  in the  following periods  of the  plasma wave  and  they are
accelerated  to lower  energies. However,  as  the load  of the  first
bucket  also damps  the field  in  the trailing  plasma buckets,  beam
loading should  prevent the trapping  of large charge after  the first
bucket and therefore reduce the dark current of the accelerator. 
\begin{figure}
\centering
\includegraphics[width=0.8\linewidth]{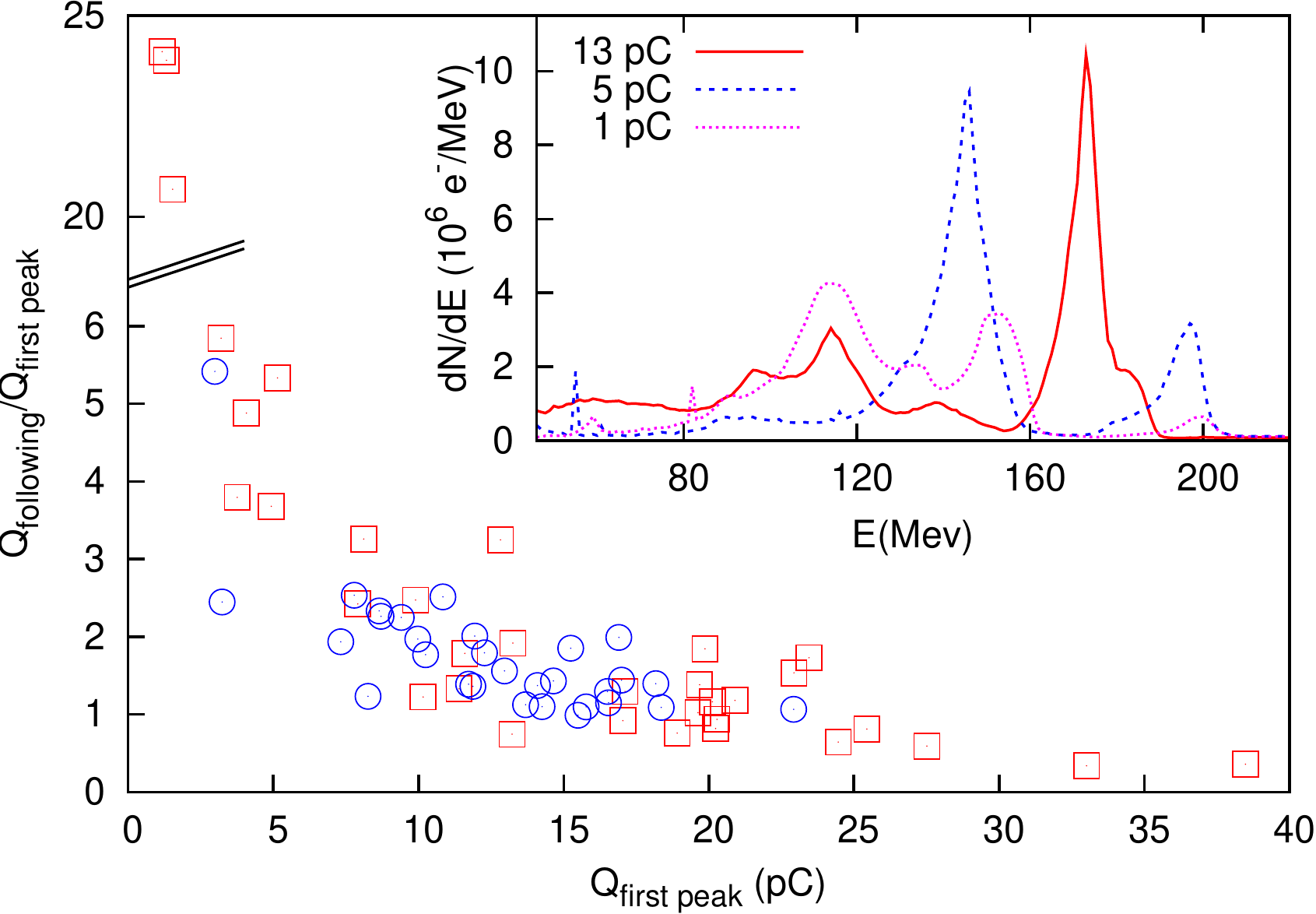}
\caption{Experimental results: Evolution of the ratio between the charge trapped in the trailing buckets (above 45 MeV) and in the first peak as a function of
 charge trapped in the first peak. Red squares correspond to the
 varying $a_1$ data set, blue circles to 30 consecutive shots with
 laser fluctuations. Inset: electron spectra for different
 charge loads. 
\label{exptr}}
\end{figure}

The inset in Fig.\ref{exptr} represents spectra with low injected
charges. Whereas for a peak charge of $13\;\mbox{pC}$ most electrons are
contained in the high energy peak, the dark current increases when the
peak charge is smaller.
Fig.\ref{exptr} represents the ratio $Q_{following}/Q_{first\,peak}$ versus
$Q_{first\,peak}$, where $Q_{first\,peak}$ is the charge in the high energy peak and
$Q_{following}$ is the charge in the rest of the distribution (above
45 MeV). This ratio represents a measurement of relative the dark current, and it clearly decreases
with the charge trapped in the first peak, as expected from the
beam loading effects.

These experimental observations reveal the effects of beam loading
but simulations are needed to fully test this interpretation and
exactly understand the role of the variation of the injection volume which
might also change the energy of the electron bunch. To
model the experiment, 3D particle in cell (PIC)
simulations have been performed with the code CALDER \cite{Calder} for similar parameters: a
normalized amplitude of $a_0=1.3$ and an electron density $n_e=7.5\times
10^{18}\; \mbox{cm}^{-3}$ for a collision position $z_{coll}=-575\;\mu\mbox{m}$ which gives a trapped charge similar to
the experimental results  \cite{Xavier}. Simulations are performed for
different values  of the  injection pulse intensity  but, to  limit the
computational    time,    the    simulations    are    stopped    only
$300\;\mu\mbox{m}$ after injection, resulting in limited acceleration,
typically to 70 MeV. 

\begin{figure}
\centering
\includegraphics[width=0.6 \linewidth]{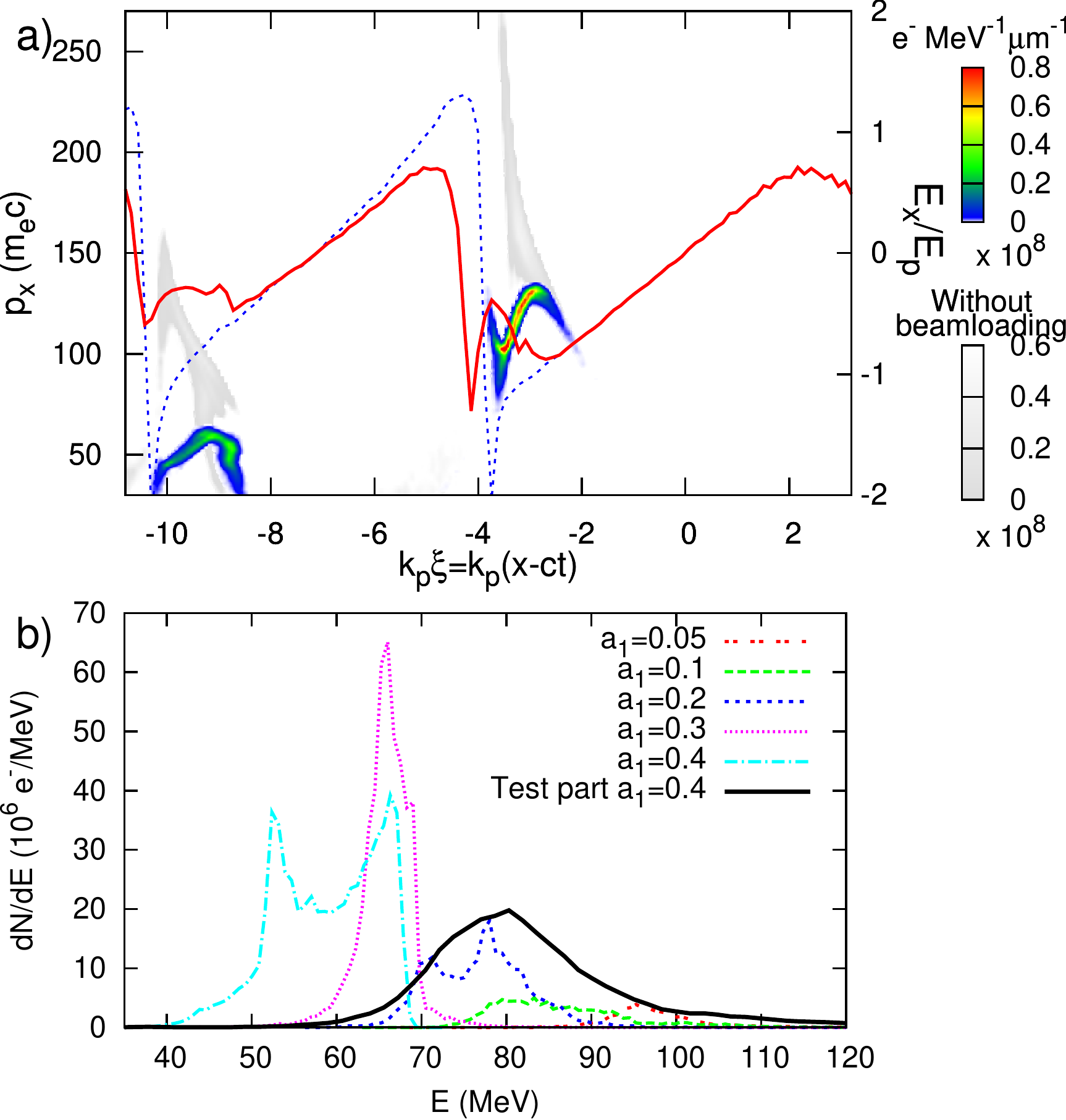}
\caption{Simulations results (Color online): a): Phase space of trapped
 electrons for $a_1=0.4$, with and without beam loading (pale gray)
 and corresponding wakefields. b): Spectra of the electrons trapped in the first bucket for
 $a_1=0.05,\,0.1,\,0.2,\,0.3,\,\mbox{and }0.4$. The thick solid line corresponds
 to the test particle case (without beam loading) with $a_1=0.4$.} 
\label{simusp}
\end{figure}

To first give a global overview of the beam loading effects, we
represent in Fig.\ref{simusp}.a the phase space of the electrons
after a $300\;\mu\mbox{m}$ acceleration for two different simulations:
both  are performed  with $a_1=0.4$  but  in the  second (pale  gray),
electrons with longitudinal momentum above $12\; m_e c$ 
are treated as test particles, i.e. they do not contribute to the plasma
fields, so that the loading of the wake is artificially removed after
injection. The longitudinal on-axis electric field is also
represented, the solid line corresponding to the loaded case and the
dotted line to the test particle case.
Whereas in the simulation without beam loading, we have the most
energetic electrons at the back of the bunch, beam
loading tends to flatten, and in this case even invert the
electric field, so that trailing electrons are heavily slowed down.
As expected, the second period of the wakefield is also damped by the loading of the
first wake.

To have a closer look on the effect of beam loading over the resulting spectra, we
represent in Fig.\ref{simusp}.b the spectra of the electrons in the
first plasma period after $300\;\mu\mbox{m}$ for five different
values of $a_1$. The figure confirms that increasing $a_1$ permits to increase the
injection volume and therefore the charge.
It also shows that the energy of the bunch decreases with the injection
pulse amplitude, as in the experiment. We first concentrate on the
small loads, with $a_1=0.05,0.1,0.2$, for which the electric field
stays monotonous over the bunch length. Two effects are combined to
explain the decrease of the bunch energy: (i) when $a_1$ and thus the injection
volume is increased, electrons can be trapped closer to the laser
pulse, in lower energy gain orbits, (ii) beam loading due to the leading 
electrons of the bunch can slow down the trailing electrons. 

To remove the ambiguity between these two effects, we also represent
the spectrum computed in the test particle simulation ($a_1=0.4$) in
Fig. \ref{simusp}.b. Comparing it to the simulation with $a_1=0.05$,
for which beam loading is also negligible, allows us to witness the influence
of a change of injection volume only. We see that the most energetic
orbits, that need the lowest initial momentum to be populated, are
loaded similarly in both cases. The only difference, linked
with the electrons gaining the highest momentum in the collision of
the lasers, lies in the low energy cut-off. 
The fact that, in the simulations and in the experiments (see inset of
Fig.\ref{expen}), the high energy cut-off of the spectra is shifted to
lower energy as the injected charge increases, is therefore a clear
signature of beam loading. 

\begin{figure}
\centering
\includegraphics[width=\linewidth]{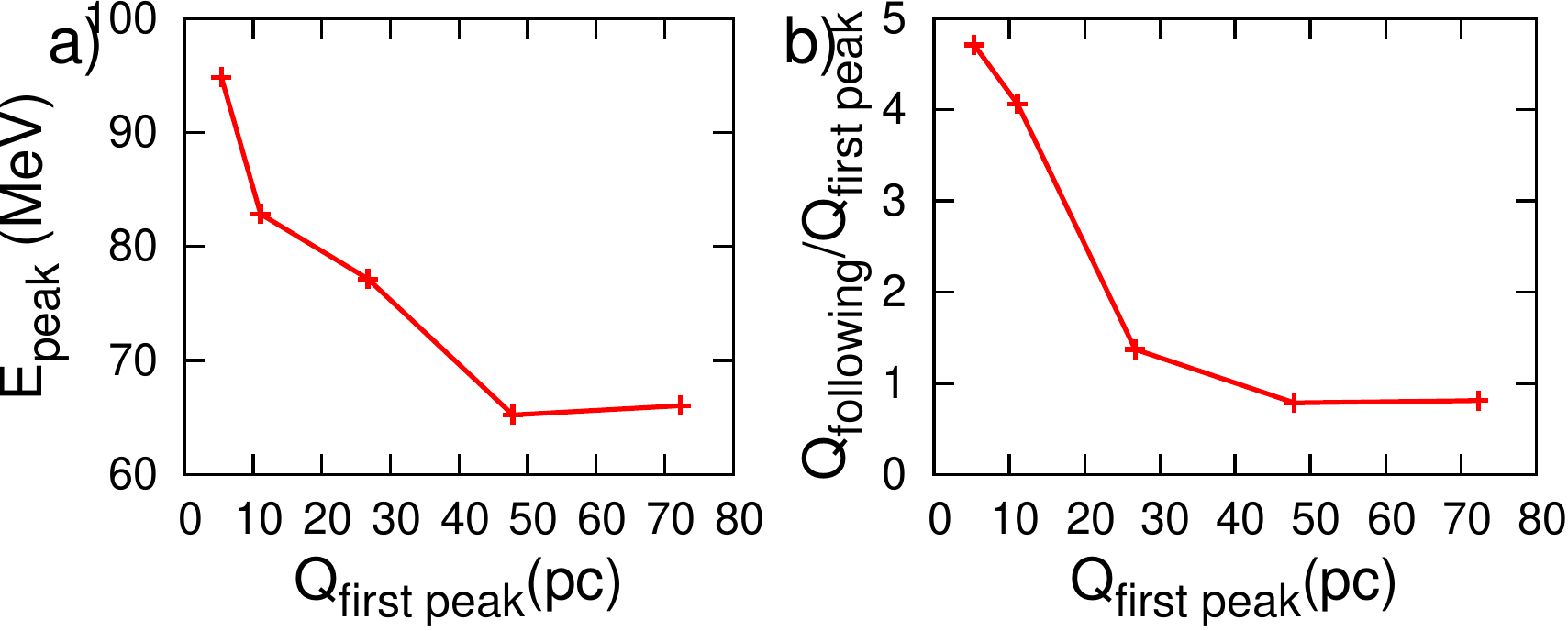}
\caption{Simulation results. a): Bunch peak energy versus charge trapped in the first bucket. b): Evolution of the ratio between the charge trapped in the trailing
 buckets (above 10 MeV) and in the first bucket as a function of
 charge trapped in the first bucket.} 
\label{simu}
\end{figure}

From the simulations, one can also deduce that the
peak energy decrease with trapped charge can be accounted
approximately for one half to the injection volume, and for the other
half to beam loading. The decrease of peak energy as a
function of charge trapped in the first bucket is represented in
Fig.\ref{simu}.a. Considering that only half of it is due to beam
loading, it gives a beam loading field per charge of approximately 1
GV/m/pC. Using the same rough estimate, the beam loading field per charge in the
experiment is close to 0.8 GV/m/pC, in good agreement with the simulations.

The simulations  with $a_1=0.3$ and  $a_1=0.4$ enlighten the
physics of beam loading for  high loads: the injected charge is indeed
so large  that it  leads to  a flattening ($a_1=0.3$)  and even  to an
inversion of the electric field ($a_1=0.4$) as shown in Fig.\ref{simusp}.a. 
When the electric  field is inverted, the trailing  electrons are less
accelerated than the leading electrons and the spectrum, showing peaks
at the electric field extrema, evolves in a different manner.
The most energetic electrons are now the leading ones and they are
not undergoing beam loading effects. Thus, the high energy cut-off and
the peak energy are now mainly determined by the injection volume, whereas
the low energy cut-off is now affected by beam loading.
This behavior explains the change of slope in Fig. \ref{simu}.a around
40  pC which  is highly  reminiscent  of the  change of  slope in  the
experimental  data  shown   in  Fig.\ref{expen}  at  approximately  20
pC. This  change of  slope occurs for  the optimal  (field flattening)
beam loading case, and the simulation case $a_1=0.3$ indeed results in
the smallest energy spread. This indicates that in our experiment, the
optimal  load  for our  accelerator  is around  20  pC.  This is  also
consistent  with  the  experimental  fact that,  for  higher  injected
charges,  structures appear  in the  first peak  and energy  spread is
rapidly increasing, see inset in Fig.\ref{expen}. 

The simulations  also show electrons  trapped in the  trailing buckets
(up to five buckets are considered due to the 
finite-size of the simulation window). Figure \ref{simu}.b represents again
the relative dark current of our accelerator versus the charge trapped in the
first bucket. The anti-correlation between  the charge in the peak and
the charge in the following  buckets is again a clear manifestation of
beam loading in the simulations, in agreement with the experiment. 

The experimental measurements presented in Fig.\ref{expen} and
Fig.\ref{exptr} have now been reproduced by simulations and they can
be interpreted as solid observations of beam loading.

In this paper, the evolution of the electron bunch
energy and dark current of the accelerator with the beam load of a well defined
wakefield are used to diagnose beam loading effects. 3D PIC simulations
show that the evolution of energy is due for one half to the variation of
the injection volume and for the other half to beam loading. This
enables us to infer an experimental beam loading field of 0.8
GV/m/pC. The evolution of the bunch energy versus charge also tells us
that the optimal load for our accelerator of about 20 pC. If we
assume a bunch duration of some fs (the simulations give a typical rms
bunch duration of 1.5 fs), this value is in good agreement with the
optimal longitudinal density of the bunch derived by Tzoufras {\em et. al} in \cite{Tzoufras}. Finally, at this charge level, the dark
current linked with the electrons  trapped in the following buckets is
also reduced. 

The implications of these observations of beam loading are crucial for
future laser-plasma accelerator designs. 
The evolution of energy spread can indeed be understood as an interplay
between injection volume and beam loading. For charges below the
optimal load (electric field not yet flattened), increasing the
injection volume will result in a larger energy spread after
acceleration and rotation in phase space, but it will also increase
the charge, resulting in field flattening and improvement of the
energy spread. The balance of the two effects will produce a
reasonably good beam quality. On the contrary, for charges above the
optimal load (inversion of the electric field), both effects will
result in an increase of energy spread, explaining the fast
deterioration of beam spectral quality. Therefore, beam loading sets
a limit on the bunch charge (tenths of pC for our 30 TW laser system),
above which the energy spread will irremediably grow. 

To further increase the charge while maintaining a good beam quality, it is
possible to use a higher laser amplitude: following \cite{Tzoufras}, in the blow
out regime, the optimal longitudinal density of the bunch scales as
$a_0^2$. But it is also mandatory to control the injection volume thoroughly
to avoid any irreducible energy spread. Current directions for downsizing the
injection volume in optical injection schemes are to use a cold
injection scheme \cite{Coldinjection} to limit the initial energy
spread. One can also reduce the plasma density: in a longer plasma period,
the injection volume, determined by the sizes the two colliding laser
pulses, would be indeed comparatively smaller. 

We acknowledge the support  of the European Community-New and Emerging
Science  and  Technology  Activity  under  the  FP6  ``Structuring  the
European  Research Area''  program (project  EuroLEAP,  contract number
028514), the support of the French National Agency ANR-05-NT05-2-41699
``ACCEL1'' and the support of RTRA through the project APPEAL.

\bibliographystyle{prl3}

\begin{thebibliography}{10}

\bibitem{TajimaPhysRevLett.43.267}
T.~Tajima and J.~M. Dawson.
\newblock Phys. Rev. Lett. \textbf{43}, 267 (1979).

\bibitem{2004Nature1}
S.~P.~D. {Mangles} et~al.
\newblock Nature \textbf{431}, 535 (2004).

\bibitem{2004Nature2}
C.~G.~R. {Geddes} et~al.
\newblock Nature \textbf{431}, 538 (2004).

\bibitem{2004Nature3}
J.~{Faure} et~al.
\newblock Nature \textbf{431}, 541 (2004).

\bibitem{LBNL1GeV}
W.~P. {Leemans} et~al.
\newblock Nature Phys. \textbf{2}, 696 (2006).

\bibitem{NatureCPI}
J.~{Faure} et~al.
\newblock Nature \textbf{444}, 737 (2006).

\bibitem{kats87}
T.~Katsouleas et~al.
\newblock Part. Acc., \textbf{22}, 81 (1987)

\bibitem{Reitsma}
A.~Reitsma, R.~Trines and V.~Goloviznin.
\newblock IEEE Trans. Plasma Sci. \textbf{28}, 1165 (2000).

\bibitem{Tzoufras}
M.~Tzoufras et~al.
\newblock Phys. Rev. Lett. \textbf{101}, 145002 (2008).

\bibitem{Esarey}
E.~Esarey et~al.
\newblock Phys. Rev. Lett. \textbf{79}, 2682 (1997).

\bibitem{Fubiani}
G.~{Fubiani}, E.~{Esarey}, C.~B. {Schroeder} and W.~P. {Leemans}.
\newblock \pre \textbf{70}, 016402 (2004).

\bibitem{PPCF}
J.~{Faure} et~al.
\newblock Plasma Phys. Control. Fusion \textbf{49}, 395 (2007).

\bibitem{PRLrec}
C.~Rechatin et~al.
\newblock Submitted to Phys. Rev. Lett. (2009).

\bibitem{aimantyannick}
Y.~{Glinec} et~al.
\newblock Rev. Sci. Instrum. \textbf{77}, 103301 (2006).

\bibitem{poprechatin}
C.~{Rechatin} et~al.
\newblock Phys. Plasmas \textbf{14}, 060702 (2007).

\bibitem{Calder}
E.~Lefebvre et~al.
\newblock Nucl. Fusion \textbf{43}, 629 (2003).

\bibitem{Xavier}
X.~Davoine et~al.
\newblock Phys. Plasmas \textbf{15}, 113102 (2008).

\bibitem{Coldinjection}
X.~Davoine et~al.
\newblock Phys. Rev. Lett. \textbf{102}, 065001 (2009).

\end{thebibliography}

\end{document}